\documentclass[fleqn,usenatbib]{mnras}
\usepackage{float}
\usepackage{amsmath}
\usepackage{amssymb}
\usepackage{rotating}
\usepackage{tabularx}
\usepackage{gensymb}
\usepackage{longtable}
\usepackage{lscape}
\usepackage{xr}

\include{journaldefs}
\usepackage[font=footnotesize,labelfont=bf]{caption}

\begin{document}

\title[\bf  Far red OB-star radial velocities ]{\bf 
Radial velocities from far red spectra of Carina Arm O and early B stars
}
\author[J. E. Drew et al]{
{\parbox{\textwidth}{J. E. Drew$^{1}$\thanks{E-mail: j.drew@ucl.ac.uk}, F. Blake-Parsons$^1$, M. Mohr-Smith$^2$}
}\\ \\
$^{1}$Department of Physics \& Astronomy, University College London, Gower Street, London, WC1E 6BT, UK \\
$^{2}$School of Physics, Astronomy \& Mathematics, University of Hertfordshire, Hatfield AL10 9AB, , UK \\
}

\maketitle

\begin{abstract}
Massive O and early B stars are important markers of recent star formation and exert a significant influence on their environments during their short lives  via photoionization and winds and when they explode as supernovae.  In the Milky Way they can be detected at great distances but often lie behind large dust columns, making detection at short wavelengths difficult.  In this study the use of the less extinguished far-red spectrum (8400 -- 8800 \AA ) for radial velocity measurement is examined. Results are reported for a sample of 164 confirmed OB stars within a 2-degree field positioned on the Carina Arm.  Most stars are at distances between 3 and 6 kpc, and Westerlund 2 is at the field edge.  The measured radial velocities have errors concentrated in the 3--10 km s$^{-1}$ range, with a systematic uncertainty of 2--3 km s$^{-1}$.  These are combined with Gaia-mission astrometry to allow full space motions to be constructed.  Up to 22 stars are likely to be runaways although 8 of them are as likely to be interloping (so far undetected) binaries.  The mean azimuthal motion of the sample fits in with recent measurements of Galactic disk rotation.  In the Galactocentric radial direction the mean motion indicates modest infall at a speed of $\sim$ 10 km s$^{-1}$. This experiment shows that weak Paschen lines in the far-red can yield credible radial velocity determination, offering the prospect of exploring OB-star kinematics over much more of the Galactic disk than has hitherto been possible. 
\end{abstract}



\section{Introduction}



Massive O and early B stars are important determinants of galactic evolution as providers of both ionizing  and bulk kinetic energy.  They are also the pre-eminent markers of recent star formation events, thanks to their short lifetimes of a few million years.  As a result of this association with recent and ongoing star formation, they frequently reside in obscured locations behind large dust columns.  The total number of these stars in the Galaxy is counted in tens of thousands rather than in the  billions generally appropriate to later-type stars, and they are commonly associated with active spiral arms at distances of a few kpc.  From the observer's point of view, these properties add up to distant, frequently highly-reddened, objects in locations we wish to understand better.

It remains the case for O and early B stars that much analysis rests on the blue spectrum for the good reason that this wavelength domain is closer to the Planck maximum in the space ultraviolet and, accordingly, is richer in photospheric absorption lines.  But if we are to build a good picture of the properties of OB populations within the major spiral arms of the Milky Way disc, it will help if we start to access a larger volume by making more use of less extinguished redder wavelengths. The first goal of this paper is to report on the feasibility of OB-star radial velocity measurement in the far-red (8400--8800 \AA ), using ground-based spectra. So far, few works focusing on early-type stars have ventured into the far-red realm, in part because not so many spectrographs are equipped with suitably-blazed moderate/high resolution gratings.  This will see a step change with the upcoming commissioning of pan-optical spectrographs for the massive-multiplex survey telescopes, offering resolutions $> 5000$ at wavelengths up to $\sim$9500~\AA\ \citep{WEAVE,4MOST}.  Interest in this wavelength domain has also been fuelled by the imminent arrival of spectra from the Radial Velocity Spectrometer \citep[RVS,][]{RVS} on board the Gaia mission.

The relative lack of attention to the far-red spectra of O and early B stars means little is in the literature.  The first comprehensive atlas that examined the far-red, or calcium triplet (CaT) region, empirically was due to \cite{Andrillat95}.  This ground-breaking work included just 6 O stars and 8 early B stars (with bright giants and supergiants making up more than half).  This has not changed much: the widely consulted MILES library of stellar spectra \citep{MILES} only contains 3 O and 7 early B stars, out of a total of 985.  As far as the most appropriate NLTE synthetic spectra are concerned, the situation is even worse for O stars in that no available grid has extended into the far-red. For B stars cooler than 30 kK the supply is appreciably better thanks to the NLTE modelling of \cite{TLUSTY}. 
Viewed in this context, the data set of 164 medium-resolution CaT-region spectra of massive OB stars hotter than 19 kK, presented here, is an order of magnitude improvement on existing empirical libraries.  

The spectra measured in this study were obtained as part of an investigation aimed at confirming the photometric selection techniques behind a new catalogue of 5915 Carina-Arm O--B2 stars \citep{MMS2017}. Their positioning within a pencil beam of just $\sim 2$ degrees diameter at distances of between 3 and 6 kpc-- not far from the Carina arm tangent -- creates an opportunity to test the strong expectation that the kiematics of these very young stars (less than $\sim10$ Myrs old) resemble that of the gas from which they formed. We find that they are, although pure circular motion is not the whole story in that the mean radial motion is convincingly non-zero.    

The contents of this paper are organised as follows.  The origin of the OB-star sample is outlined in more detail Section~\ref{sec:sample}, and is followed in Section~\ref{sec:spectroscopy} by a description of the spectroscopic data 
 The criteria for inclusion in the sample measured for radial velocity are set down in Section~\ref{sec:sample_criteria}.  Our analysis, including an overview of the available absorption lines within the range and a test of the radial velocity registration, is provided in Section~\ref{sec:analysis}.
 The presentation of our results is in Section~\ref{sec:results}, where the measured radial velocities are augmented with proper motions from the Gaia mission in order to obtain full space velocities.  We identify the likely runaway stars in the sample and describe the emergent average motion in all 3 components in the Galactocentric frame.  There is then an evaluative discussion (Section~\ref{sec:discussion}) in which: notes are presented on runaways not ejected from Westerlund 2; the impact of binarity on the dispersion of the radial velocities is considered; comparisons are made between the mean Galactocentric motion of the OB sample with those of young open clusters (including Westerlund 2) and gas tracers. The paper ends with our conclusions (Section~\ref{sec:conclusions}). 

\section{Data}

\subsection{The OB-star sample}
\label{sec:sample}

The work presented here builds on two previous studies.  The first of them by \cite{MMS2017} provided both a catalogue of photometrically-selected O and early B stars in Carina, and a spectroscopic test of the quality of selection. Since then, the catalogue has been put to use in examining OB stars around the well-known massive young clusters Westerlund 2 and NGC 3603 \citep{Drew2018, Drew2019}.  It has also been the basis for a proper motion study of a much larger $4\times12$ sq.deg sky region capturing the far Carina Arm \cite{Drew21}. Here, we return to the spectroscopic observations, obtained with AAOmega at the Australian Astronomical Telescope (AAT) in 2014, in order to present the so-far unpublished far-red spectra obtained at the same time as the blue spectra already analysed for effective temperature and surface gravity \cite{MMS2017}. Our focus, now, is on examining the utility of moderate spectral-resolution far-red spectra for radial velocity measurement. 

\begin{figure}
\begin{center}
\includegraphics[width=1.0\columnwidth]{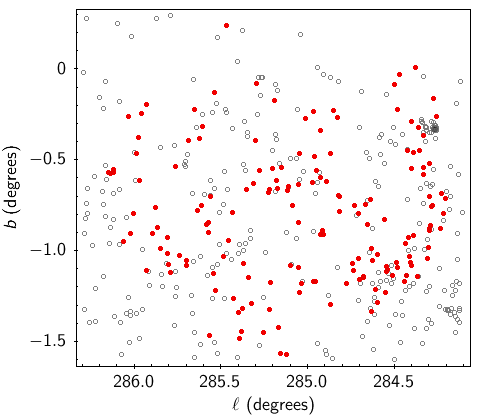}
\caption{
The sky positions of the 164 stars with spectra adequate for radial velocity analysis (red points), along with the positions of stars contained in the larger sample of Drew et al. (2021) in their neighbourhood (grey circles).  The massive young cluster, Westerlund 2, is clearly visible near $\ell = 284^{\circ}.3$, $b = -0^{\circ}.3$.
}
\label{fig:map}
\end{center}
\end{figure}

The \cite{MMS2017} spectroscopic sample contains $\sim$190 non-emission OB stars, of which 164, ultimately, have well-enough exposed spectra to support radial velocity measurement. Westerlund 2 sits on the edge of the sampled region.  The exact selection cuts applied in choosing the spectroscopic subset are set out in Section~\ref{sec:sample_criteria}.   Almost all of the final sample selected for analysis appeared in the proper-motion selection of 4199 OB stars constructed by \cite{Drew21}.  Figure~\ref{fig:map} shows the sky positions of these objects, along with the positions of the 435 stars from the larger \cite{Drew21} list in the same area.  

\begin{figure}
\begin{center}
\includegraphics[width=1.0\columnwidth]{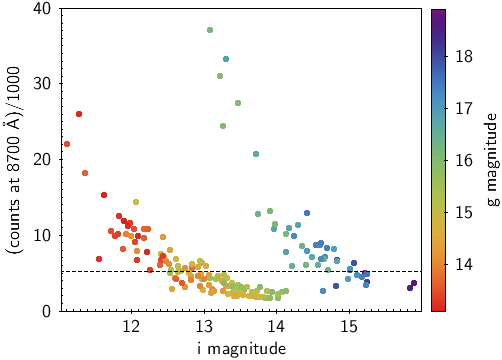}
\caption{Counts at 8700 \AA\ in 1000s in the 1700D spectra analysed, as a function of $i$ magnitude (from the MS17 catalogue).  The data points have been coloured according to the $g$ magnitude, which determined the total AAOmega exposure time as explained in the text. The median count level of 5200 is marked by the horizontal black dashed line.
}
\label{fig:counts}
\end{center}
\end{figure} 


\subsection{Spectroscopic observations}
\label{sec:spectroscopy}

A first account of the data collection at the AAT on 18th June and 3rd July 2014 was provided by \cite{MMS2017}.  A key point relevant to this study is that the target selection was geared towards secure detection in the blue and that two 2dF/AAOmega configurations were observed: the brighter, with $13 < g < 16$ targets being exposed for 40 minutes once on 18th June, while fainter stars with $16 < g < 19$ were exposed (and later co-added) for a total of 140 minutes over both nights.  The slit width was 0.5 arcsec.  The red arm grating installed was the 1700D grating which offers a free spectral range running from 8340~\AA\ to 8820~\AA\ in most fibres, and a resolving power of around 11000, at the longer wavelengths in the range. 

The spread in interstellar extinction affecting the stars observed, combined with the change in exposure time at $g = 16$, produces a non-monotonic relationship between observed counts and any red magnitude. This is demonstrated in Fig.~\ref{fig:counts}, in which sky-subtracted counts measured in a 40~\AA\ band centered on 8700~\AA\ are plotted against $i$ magnitude.  We have not made use of measurements on spectra with fewer than $\sim$1650 sky-subtracted counts per pixel -- a level that delivers a signal-to-noise ratio of 28. The $g$ magnitudes of the two objects at this bottom end of the range have $g$ magnitudes of 15.86 and 15.97.  The median count at 8700 \AA\ for the final sample is 5200 counts, corresponding to a signal-to-noise ratio of $\sim$50.

The data were extracted using the {\sc 2dfdr} package.  The sky subtraction made use of dedicated sky fibres.  The final wavelength scale applied to the spectra is in the lab frame.  For the epoch of observation the corrections to the heliocentric frame and to the local standard of rest are respectively blueshifts of $\sim 14$ and $\sim 24$ km s$^{-1}$. 

%

\subsection{Criteria for inclusion in the analysed sample, and distances}
\label{sec:sample_criteria}

The cuts applied to the full set of stars with spectra pay attention both to astrometric and spectroscopic and properties. 

The astrometric requirements are that a Gaia EDR3 parallax and proper motion should be available and that the these data are reliable (RUWE $< 1.4$).  It is also required that the ratio of the parallax measured to its error, ($\pi/\sigma_{\pi}$, should exceed 3.  The aim of this criterion is to achieve a compromise between limiting the impact of distance error and the retention of a good-sized sample.  

The spectroscopic requirements are as follows:-
\begin{itemize}

    \item There should be no Paschen line emission in the far red spectrum.
    \item As noted already above, the minimum acceptable count level at 8700 \AA\ is 1650 per pixel.  (178 spectra/stars meet these first two criteria, along with the astrometric requirements.) 
    \item The available estimate for $\log{T_{{\rm eff}}}$ should exceed 4.28. For all but 10 percent of the stars the effective temperature measures are taken from the analysis of the blue spectrum \citep{MMS2017}. This threshold, corresponding to 19 kK, was set in order to remove stars at risk of undetected calcium infrared triplet/Paschen line blends.  It was imposed following on from trials of radial velocity measurement that showed stars cooler than this threshold gave different results depending on whether the calcium lines were assumed to be present or not. This ambiguity vanishes in hotter stars.
    \item Three further spectra, proved too difficult to measure for radial velocity with any confidence.  Two of them were at the bottom end of the counts range and very noisy.  The third (\#1528) was well exposed but clearly a spectroscopic binary and so unsuitable for present purposes.  These were also left out. 
\end{itemize}

The list remaining after these exclusions comprises 164 stars with spectra. A table identifying and summarising their parameters and measurements made is given in supplementary materials.  Appendix~\ref{sec:full_list} describes the content of the table.  46 of the stars have effective temperatures exceeding 25 kK, leaving 118 in the range 19--25 kK.   Inevitably, there is a strong weighting to the bottom part of the accepted temperature range.  There will be cause, in Section~\ref{sec:rv_check}, to pick up four spectra not in this main list as examples of superposed narrow (i.e. nebular) Paschen line emission.  Measurements of the emission-line radial velocities in these instances can be compared with literature measures as a useful check on the wavelength scale.

Distances are inferred from the Gaia EDR3 parallaxes in exactly the same way as by \cite{Drew21}.  This is application of the EDSD method described by \cite{Luri2018}, for which an estimate of a length scale over which the stellar density declines is required.  It was set to be 2kpc for the reasons set out by \cite{Drew21}.  Again, for the reasons given by \cite{Drew21}, a global parallax offset of 30 $\mu$arcsec was assumed.

The availability of Gaia astrometry is crucial to the constructed sample.  The recent DR3 release has added a new wealth of spectroscopically-based stellar parameters, including radial velocities for some stars.  A cross-match reveals that just 12 of the 164 stars now have Gaia radial velocities.  Unfortunately, these have to be discarded as unreliable because of the combined impact of serious errors in the derived atmospheric parameters and the pipeline reliance on spectral template comparisons (see Appendix~\ref{sec:Gaia_spectra} for more detail).  Accordingly, we continue to exploit just the astrometry in this study.


\section{The OB-star far red spectra and their analysis}
\label{sec:analysis}

\begin{figure*}
\begin{center}
\includegraphics[
width=2.0\columnwidth]{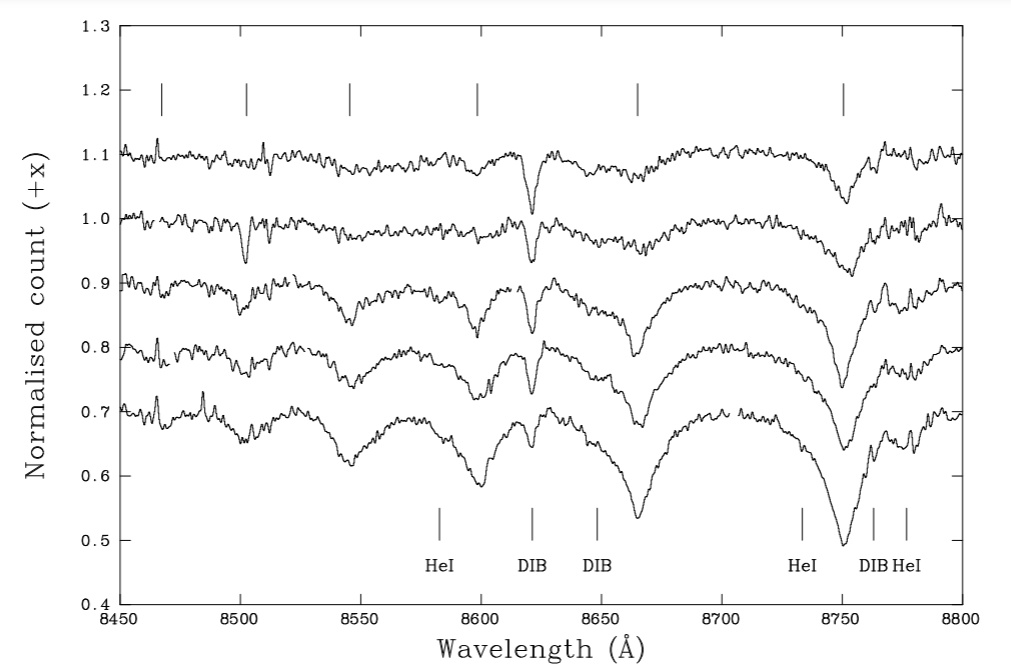}
\caption{Normalised spectra spanning the effective temperature range of the sample.  The wavelength range is limited 350 \AA\, running from 8450 to 8800 \AA , in order to better display the readily detected absoprtion features.  From bottom to top the stars are: \#3207 (20.2 kK), \#2497 (23.2 kK), \#2662 (27.8 kK), \#2164 (34.9 kK) and \#1338 (42.5 kK).  All have gravity estimates appropriate for dwarfs.  The positions of the Paschen lines (running from Pa12 at 8750.47 \AA\ to Pa17 at 8467.25 \AA ) are marked across the top, while the HeI and DIB line wavelengths are marked below.
}   
\label{fig:spectral_templates}
\end{center}
\end{figure*}

\subsection{Overview of the far red spectrum}

As so little attention has been paid, hitherto, to the far-red spectrum of OB stars, it is helpful to consider the spectral features within it before proceeding to radial velocity measurement.  We show in Figure~\ref{fig:spectral_templates} five well-exposed 1700D spectra that span the effective temperature range of the sample, from 20 kK up to 42.5 kK.  

Paschen series lines, along with the 8621~\AA\ diffuse interstellar band (DIB), are the most obvious absorption lines present: the rest wavelengths of P12 up to P18 fall in the observed range. They are easy to pick out up to P16 at around $T_{{\rm eff}} \sim 20$~kK, but in the earliest O stars it can be hard to make out even P13. At effective temperatures up to 19 kK, the Ca{\sc ii} IR triplet components blend with P13, P15 and P16 and will blueshift measures of these lines' central wavelengths if not accounted for, even when the Ca{\sc ii} lines are not obvious to the eye.  The measurement challenge posed by this part of the spectrum arises from the combination of relatively weak stellar lines (mostly dropping no further than 10\% below continuum) and the noise introduced by sky-line residuals.    

\cite{Andrillat95} identified two lines of He{\sc i} in the range at 8582.7 and 8733.4 \AA .  A third, He{\sc i} line, just beyond the range of their atlas, is seen in our data at 8776.7 \AA .  It is the strongest of the three and it notably disappears, along with any sign of the other two lines, in the O-star range (see the top two spectra in Figure~\ref{fig:spectral_templates}).  This line coincides with some sky lines that frequently leave residuals in the sky-subtracted spectrum.

The DIB at 8621~\AA\ would be expected to be pronounced in the spectra, given the distances mostly exceed 3~kpc and the interstellar column is boosted by the passage of the line of sight through the near Carina Arm.  The median distance and extinction for the final sample analysed here are respectively 4.6~kpc and $A_0 = 4.0$~mag.  \cite{Jenniskens94} and \cite{Herbig95} agree on a broader and shallower DIB in the range of our spectra at $\sim$8648~\AA .  However, it does not always show well, sitting as it does in a difficult part of the spectrum prone to low-level changes from object to object - most likely made worse by troubles tracing and subtracting a band of sky emission here.  In Fig.~\ref{fig:spectral_templates}, it is most obvious in the spectrum of \#1338, at the top.  \cite{Herbig95} also mentions a possible DIB at $\sim$8763~\AA .  This is sometimes detected in the red wing of Pa12. Along with the already-mentioned sky emission here and the HeI line, it interferes with the smooth return of Pa12's red wing to the continuum.  

\label{sec:properties}


\subsection{Radial velocity determination}
\label{sec:rv_measures}

\begin{table}
\caption{Wavelength limits of the windows used to define the quadratic continuum in each spectrum.  The total number of spectral pixels captured within the windows is 156.
}
{\centering
\begin{tabular}{c}
\hline
Window wavelength limits (\AA )\\
\hline
  8420.1 -- 8428.1 \\
  8480.9 -- 8486.5 \\
  8520.8 -- 8527.2 \\
  8626.3 -- 8628.7 \\
  8701.4 -- 8709.4 \\
  8798.9 -- 8806.9 \\
\hline  
\end{tabular}
 }
\label{tab:continuum_windows} 
\end{table}

The measurement of radial velocity (RV) in the available far-red spectra is challenging, thanks mainly to the low contrast of the Paschen lines and their breadth.  The high but very different extinctions (ranging from 2 up to 8 visual magnitudes) are also a potential issue although it helps that they have been determined to sufficient precision for all objects from SED fitting \citep[see][]{MMS2015, MMS2017}.  Because, at the outset, it was not obvious how best to proceed with RV measurement, a number of trials were undertaken varying the number of lines used with the aim of finding the best method that could be applied uniformly.  After trials, we settled on fitting simple Gaussians to the cores of just the Pa12 and Pa13 lines on the grounds that these lines are nearly always both apparent, while sometimes affected by blending in their wings with other features.  Indeed in stars cooler than $\sim$ 19kK (omitted, see Section~\ref{sec:sample_criteria}), a line of the calcium triplet is capable of blue-shifting the apparent mean wavelength of Pa13.  As very low optical depth high members of the Paschen series, forming deep in the photosphere, Pa12 and Pa13 should experience negligible profile modification by mass loss.  The alternative technique of cross-correlation, or indeed, the fitting of appropriately broadened templates offers no advantage presently, given the small number of uncompromised photospheric features available to apply them to.  Our simple approach captures the required information.

The spectra were prepared for measurement as follows.  Every spectrum was dereddened on the basis of the object's SED fit, using a mean Galactic extinction law (due to \cite{Howarth83} in this instance), and renormalised to 10\,000 counts at 8700 \AA .  It was then possible to check that the overall continuum shape stayed broadly constant from spectrum to spectrum.  This was generally the case and it made it easier to apply a uniform method of continuum fitting. The bowed shape present in all the spectra was fit to a quadratic, adopting the wavelength ranges set out in Table~\ref{tab:continuum_windows} as continuum windows. There were a few spectra for which it was necessary to limit the longest wavelength window to a shorter wavelength range (still redward of Pa12) to deal with the occasional spectral range truncation.  

\begin{figure}
\begin{center}
\includegraphics[width=1.0\columnwidth]{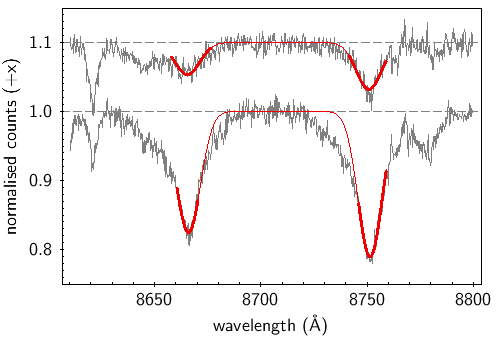}
\caption{Two contrasting examples of the Pa12 + Pa 13 Gaussian fits used to determine radial velocity.  The top spectrum is of a hot O star with weak lines (\#1338, $T_{{\rm eff}} = 42.5$ kK), while the bottom is one of the coolest exhibiting maximum line contrast (\#2179, $T_{{\rm eff}} = 19.5$ kK).  Both are shown in grey.  The fits returned are overplotted in red, using bold style to pick out the limited wavelength ranges passed to the fitting software. The fit errors in these cases are respectively 6.5 and 2.1 km s$^{-1}$ (i.e. median and best). 
}
\label{fig:line_fits}
\end{center}
\end{figure} 

\begin{figure}
\begin{center}
\includegraphics[width=1.0\columnwidth]{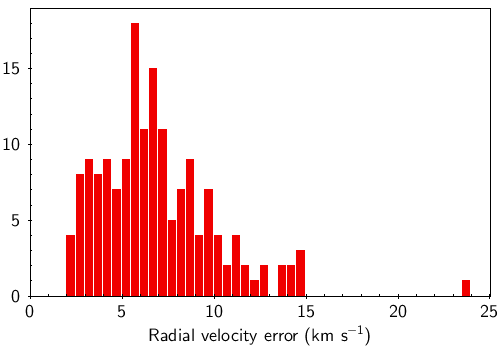}
\caption{Histogram of radial velocity measurement errors. 
}
\label{fig:rv_errors}
\end{center}
\end{figure}

Once each spectrum had been normalised, the core regions of the Pa12 and Pa 13 lines were identified and fit to two Gaussians under the constraint that the wavelength difference between the fit line centres is fixed at 84.454 ~\AA  -- the difference between the Pa12 and Pa13 rest wavelengths.  The widths of the Gaussians were also required to be the same. The fitting windows placed around each of the lines were set interactively (depending on apparent line breadth and, from time to time, noise features): they ranged from 10 to 20 \AA\ wide.   Examples of how this worked out are presented for two examples taken from opposite ends of the effective temperature range in Fig.~\ref{fig:line_fits}.  The software used to perform the fits within the {\sc DIPSO} analysis package are the {\sc ELF} routines\footnote{These are designed for emission line fitting (hence ELF) but can be applied to absorption profiles by the simple device of subtracting the unit continuum and multiplying the negative spectrum by $-1$.}. The software returns an error estimate for the best-fit central wavelength, simultaneously with errors on line width and total area, by tracing the behaviour of the $\chi^2$ surface around an iteratively-determined best-fit minimum.  The particular method used to achieve this is the parabolic expansion of $\chi^2$ as described by \cite[][see section 8.5]{BR2003}  Expressed as velocity shifts, the errors start at 2.1 km s$^{-1}$, pass through a median of 6.5 km s$^{-1}$, and reach 24 km s$^{-1}$ in an outlying worst case.  The full distribution is shown in Fig.~\ref{fig:rv_errors}.

\begin{table*}
\caption{Radial velocities of the narrow nebular emission seen in the spectra of stars on the NE edge of RCW 49.  Both heliocentric and LSR velocities are given, as well as the lab frame, since the former are appropriate for comparison with Zeidler et al 2018 and the latter apply when comparing with Caswell \& Haynes (1987).  The errors are the same on all three frames and, hence, are not repeated.
}
{\centering
\begin{tabular}{lccccc}
\hline
Object & \multicolumn{2}{c}{Line RV} & \multicolumn{3}{c}{Weighted mean RV} \\ 
Cat. \#  &  Pa12  & Pa13  & Lab frame & heliocentric & LSR \\
  & \multicolumn{5}{c}{km s$^{-1}$} \\
\hline
1313 & 21.4$\pm$0.9 & 20.8$\pm$1.1 & 21.1$\pm$0.7 & 6.7 & $-$5.5 \\
1314 & 26.6$\pm$1.1 & 26.5$\pm$0.5 & 26.5$\pm$0.5 & 12.1 & $-$0.1 \\
1318 & 27.1$\pm$1.1 & 25.7$\pm$0.5 & 26.1$\pm$0.5 & 12.2 & 0.0 \\
1323 & 35.5$\pm$0.7 & 37.6$\pm$1.8 & 36.2$\pm$0.7 & 21.8 & 9.6 \\
\hline  
\end{tabular}
 }
\label{tab:neb_lines} 
\end{table*}

The next step is to assess the validity of the derived velocities.

\subsection{Tests of the radial velocity scale}
\label{sec:rv_check}

The observations obtained did not include radial velocity standards, leaving open the question of the reliability of the achieved radial velocity scale.  As compensation, the data obtained allow two distinct checks to be made.  

First, the stars \#1273 and \#1338 have both been observed with the VLT's X-Shooter instrument on several occasions. Radial velocities were measured from these data by \cite{Drew2018}. Binary reflex motion was not detected in either case.  Their heliocentric velocities were determined to be $+16$ and $+1$ km s$^{-1}$ respectively, relative to O stars in Westerlund 2 measured by \cite{Rauw2011}.  The uncertainties on these values are in the region of 2 km s$^{-1}$.  Our measurements of the 1700D spectra for these two stars are $+15.5 \pm 7.1$ km s$^{-1}$ and $-2.2 \pm 6.5$ km s$^{-1}$.  To well within the errors they are a match, although the new velocities for both stars are a shade more negative.

The second check available involves the spectra of four stars outside the main sample: they were not included because there are prominent narrow nebular emission superposed on the Paschen series absorption profiles.  These stars, \#1313, \#1314, \#1318 and \#1323, are positioned around 5 arcmin east of Westerlund 2 and fall within a patch of nebulosity on the edge of the RCW 49 HII region.  In these instances, radial velocities can be measured from the bright nebular Pa 12 and Pa13 emission, and be compared with measures from both radio recombination line data \citep{CandH87} and the recent optical work of \cite{Zeidler18}.  The measurements for the four stars are set out in Table~\ref{tab:neb_lines}.

\cite{CandH87} give the position and LSR radial velocity of RCW 49 as $\ell = 284^{\circ}.308$, $b = -0^{\circ}.334$ and 0 km s$^{-1}$.  The size of the radio-bright H{\sc ii} region is stated to be $5\times$7 arcmin.  The four stars with bright measurable nebular emission lie north-east of the radio position at separations of between 3 and 4 arcmin.  The specification of the Sun's motion relative to the LSR applied by \cite{CandH87} is reused here, to achieve consistency.  The mean of the 4 Paschen-emission velocities, converted to LSR and given in the final column of  Table~\ref{tab:neb_lines}, is 1 km s$^{-1}$. This is satisfactory agreement in the presence of errors on the order of 1 km s$^{-1}$.  However, it is striking that the four individual pairs of measurements are quite dispersed.  Why this might be so is clarified by optical observations of the immediate environment of Westerlund 2 by \cite{Zeidler18}, these included some measurements of nebular emission\footnote{Their measurements are reported in the barycentric frame, as delivered by the MUSE instrument's pipeline.  This differs from the heliocentric frame at a level too small to matter here.}.  Of the six RVs they report, five range between 8.2 and 11.5 km s$^{-1}$ and the sixth is 23.1 km s$^{-1}$.  This is echoed in their result of a bimodal distribution of cluster-star velocities with the larger peak at 8.1 and a second peak at 25.4 km s$^{-1}$.  The four measures in column 5 of Table~\ref{tab:neb_lines} fall in with this pattern: three of the four fall in the range 6.7 to 12.2 km s$^{-1}$, whilst the fourth is 21.8 km s$^{-1}$.  A perfect match is not to be expected given that the sky locations sampled are $\sim$5~arcmin apart.  

Put together, these comparisons give no cause to introduce a systematic adjustment to the radial velocities measured for the main sample of stars.  If there is a systematic difference it is unlikely to be more than 2--3 km s$^{-1}$ -- or no larger than the lowest random errors obtained (Fig.~\ref{fig:rv_errors}).  



\section{Results}
\label{sec:results}

\subsection{The radial velocities}
\label{sec:rv}

\begin{figure}
\begin{center}
\includegraphics[width=1.0\columnwidth]{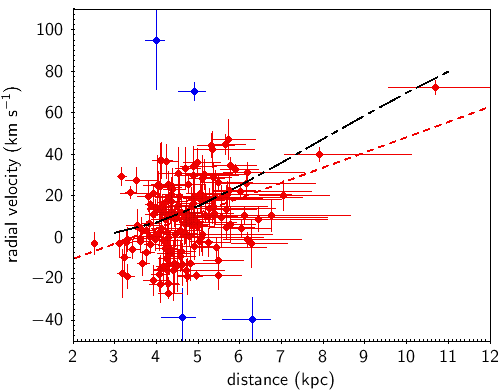}
\caption{Radial velocity as a function of inferred distance. The obvious outlier stars are picked out in blue.  The remaining sample of 160 stars, shown in red, give rise to the linear regression plotted as a dashed red line. The black dash-dotted line is the trend expected for pure rotation at a fixed circular speed of 236 km s$^{-1}$, with no inward or outward radial motion. Section~\ref{sec:space_vel} and Fig.~\ref{fig:rv_v_distance} provide the motivation for choosing 236 km s$^{-1}$.
}
\label{fig:rv_v_distance}
\end{center}
\end{figure} 

\begin{figure}
\begin{center}
\includegraphics[width=1.0\columnwidth]{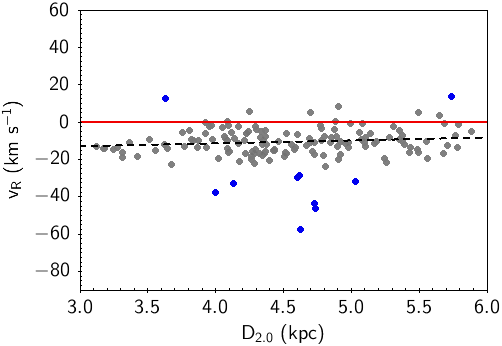}
\includegraphics[width=1.0\columnwidth]{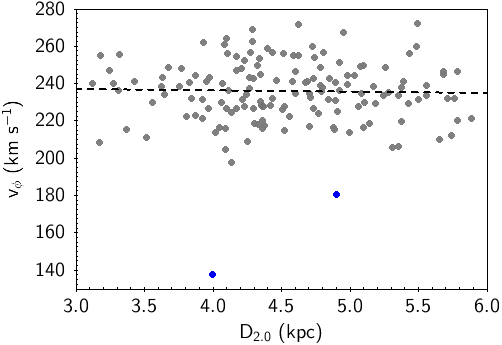}
\includegraphics[width=1.0\columnwidth]{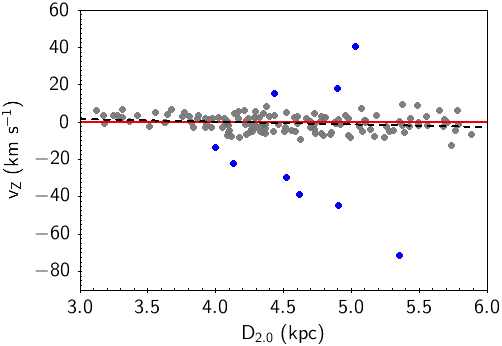}
\caption
{Trends in OB-star $v_R$, $v_{\phi}$ and $v_Z$ with distance in the best-populated 3 to 6 kpc distance range.  The velocity range in each panel is 150 km s$^{-1}$ for ease of comparison.  The linear regression in each case is drawn as a black-dashed line.  The data points in blue were left out of the fit, as obvious outliers.  The $v_{\phi}$ component is most closely aligned, in space, with the observed RV and hence shares much of its scatter.  The regression line indicates just a drop of 2 km s$^{-1}$ in $v_{\phi}$ from $D_{2.0} = 3$ to 6 kpc.  The trends in $v_R$ and $v_Z$ are more pronounced, but still modest.  Negative non-zero $V_R$ is strongly favoured, as demonstrated by the offset of the data from the zero line picked out in red.  
}
\label{fig:uvw_v_RG}
\end{center}
\end{figure}

The heliocentric RV obtained for the sample of 164 stars is shown in Fig.~\ref{fig:rv_v_distance}.  In deriving these results, the solar motion in the radial and vertical dimensions is taken from \cite{Schoenrich2010}, while the total circular motion of the Sun (summing rotation and the Sun's peculiar motion is adapted from the result of \cite{Reid2019} using the \cite{Abuter19} measurement of the radius of the Sun's orbit.  The three components of solar motion, ($v_R$, $v_\phi$, $v_Z$), are $(11.1, 248, 7.25)$ km s$^{-1}$. 

It is clear from the plot that almost all the stars lie in the distance range from 3 to $\sim$6 kpc.  Only four objects have RVs that stand apart from the main distribution (in blue, in Fig.~\ref{fig:rv_v_distance}),  If the sample is trimmed by removing the RV outliers, the best-fit linear regression is as shown (red dashed line).  If the sample is further improved by limiting the distance range to the $3 < D_{2.0} < 6$~kpc range, and by demanding $\pi/\sigma_\pi > 10$ -- reducing the number of objects to 104 -- the trend steepens very slightly (its gradient changes by less than 1\%).  The apparent upward rise is robust, and to be expected as shall be confirmed in Section~\ref{sec:space_vel}.  In advance of that, Figure~\ref{fig:rv_v_distance} also includes the trend given by a Galactic rotation law in which the circular speed is independent of Galactocentric radius and fixed at 236 km s$^{-1}$ (black dashed-dotted line).

A feature of the distribution is that the scatter about the mean trend is greater than would be expected from measurement error alone.  For the red data points shown in Fig.~\ref{fig:rv_v_distance}, the 1-$\sigma$ scatter about the trend line is 15 km s$^{-1}$.  It remains stubbornly close to this, even as the requirements on $\pi/\sigma_\pi$ and RV error are tightened up: for instance, if an upper limit is placed on RV error equivalent to the median for the sample of 6.5 km s$^{-1}$ and $\pi/\sigma > 10$ is required (55 stars), the 1-$\sigma$ scatter only drops to 14 km s$^{-1}$.  Accordingly much of it must have an astronomical origin.  A plausible contributor to the scatter would be unrecognised displacements away from the systemic RV by binary motion.  This will broaden the distribution without biasing it.  There is a discussion of this issue in Section~\ref{sec:binary_motion}.

\subsection{Space velocities}
\label{sec:space_vel}

With the astrometry in hand along with the RV measurements, the components of motion in the Galactic frame have been determined using the relations laid out by \cite{Harris19} as their equations 1 and 2 (for $u, v, w$, where it should be understood $u = - v_R$, $v = v_{\phi}$ and $w = v_\phi$).  
A step in this process involves the calculation of the Galactocentric radii of the stars in the sample.  It is useful to note that these run only from $\sim7.9$ kpc at around $\sim$3 kpc, up to just $\sim8.7$ kpc at 6 kpc.  In effect, the pencil beam occupied by our spectroscopic sample lies at a small angle relative to the Solar Circle.   This prompts the expectation that the $v_{\phi}$ component of motion should not show much variation as a function of inferred distance, given that the Galactocentric radius changes by less than 1 kpc. For example, the rotation law of \cite{Eilers19} predicts a decline rate of 1.7$\pm$0.1 km s$^{-1}$ kpc$^{-1}$.  

Our discussion now focuses on the 3 to 6 kpc distance range as this is well-populated,  offering useful statistics.  Fig.~\ref{fig:uvw_v_RG} presents the dependence of $v_R$, $v_{\phi}$, and $v_Z$ 
within these limits.  It is indeed the case that $v_{\phi}$, the azimuthal component, shows at best very little change across the range.  A formal regression applied to all but two of the stars (left out as obvious outliers) suggests a drift down from 237 km s$^{-1}$ at 3 kpc to 235 km s$^{-1}$ at 6 kpc.  The dispersion, at $\sigma \sim 15$ km s$^{-1}$ around the regression line, is considerable.  This is inherited largely from the measured radial velocities thanks to the near alignment of the $v_{\phi}$ component and the line of sight -- its median value for the sample is 235.9 km s$^{-1}$.  On excluding the two obvious outliers, this nudges up to 236.2 km s$^{-1}$.  The formal error on this is $\pm$1.2 km s$^{-1}$. The plot of observed radial velocities in Fig.~\ref{fig:uvw_v_RG} includes a curve for a constant circular rotation at 236 km s${-1}$: the small positive shift of a few km s$^{-1}$ of this curve relative to the mean trend stems from the small component of radial inflow, considered next.

The $v_R$ and $v_Z$ components are less scattered, with dispersions of $\sim 7$ and $\sim 4$ km s$^{-1}$ respectively.  In both cases the trend plots include a a larger number of obvious outliers than are seen in $v_{\phi}$.  On leaving them out, the computed linear trends again have modest negative slopes.  In the case of $v_R$, the implied rise from 3 to 6 kpc is from $-13.2$ to $-8.6$ km s$^{-1}$.  The most important feature, that is entirely robust against trimming or attempting to improve the sample (cf. Section~\ref{sec:rv_measures}), is that the values of $v_R$ are overwhelmingly non-zero and negative: the median for the dataset, less the outliers (in blue in Fig.~\ref{fig:uvw_v_RG}) is $-11.2$ km s$^{-1}$, implying a net motion towards the Galactic centre for the OB stars in the region. 

Systematic error in the radial velocity measurements is extremely unlikely to have produced this pronounced negative $v_R$ drift.  To cancel it out, they would have to be increased by $~$30 km s$^{-1}$, which would then drag the inferred circular speeds averaging down to between 200 and 210 km s$^{-1}$ (relatively low values in the context of recent work -- see Section 5.3).  Our estimate of the systematic uncertainty is only 2--3 km s$^{-1}$ (Section~\ref{sec:rv_check}). Furthermore, the proper motion data are certainly secure: across a field of $\sim2^{\circ}$ diameter, the covariance is likely to be $< 0.02$ mas yr$^{-1}$ \citep[see equations 25 and 27 in][]{Lindegren21a}, translating to under 1 km s$^{-1}$ here.

In contrast, the $v_Z$ component distribution offers no surprises, other than, perhaps, the presence of the largest number of outlier stars.  Dynamically-cold young OB stars should show no significant motion out of the Galactic plane, in bulk, and that appears to be the case.  The median $v_Z$ component is $-0.6$ km s$^{-1}$.

%



\subsection{The outlier objects}
\label{sec:extreme}

\begin{figure}
\begin{center}
\includegraphics[width=1.0\columnwidth]{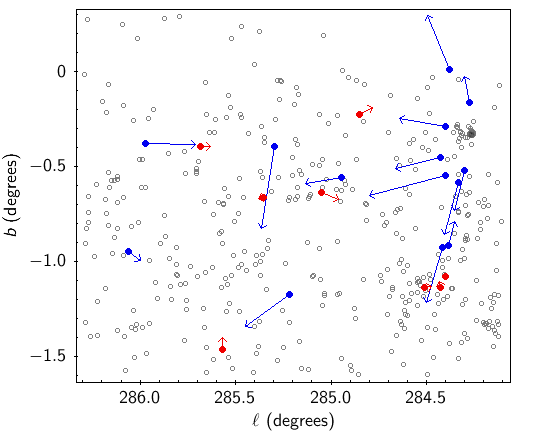}
\caption{Sky positions and relative proper motion vectors for the stars with outlying space velocities listed in Table~\ref{tab:extreme}.  The proper motion vectors plotted are proportional to the difference between their Gaia EDR3 measured proper motions and the mean for the 150 stars with spectra in the distance range $3 < D_{2.0} < 6$ kpc. Stars in red (rather than blue) are the lower proper motion objects that raise the suspicion they may be spectroscopic binaries. Stars in grey are other OB stars in the region \citep[from the list of][]{Drew21}.
}
\label{fig:outlier_map}
\end{center}
\end{figure}

We make use of two alternative criteria to determine which stars of the 150 in the $3 < D_{2.0} < 6$~kpc range are 'outliers'.
First, we collect together the objects plotted in blue in each panel of Fig.~\ref{fig:uvw_v_RG}.  These stars are deemed outliers on the basis that they are more than $3\sigma$ from the distribution mean.  Some objects meet the criterion in more than one velocity component, with the result that 15 objects are responsible for the 19 data points.  Just one object, in the MS17 catalogue as \#1356, is an outlier in all 3 velocity components: it has the largest measured radial velocity (and error) in the sample (see Table~\ref{tab:extreme}).  The nature of this object is assessed from the data available in the discussion (Section~\ref{sec:high_vel}).   It is striking that the large inherent scatter in $v$ has the consequence that only two stars sit outside the main run of the $v$ component distribution. 

The second approach we take is to compute full 3D space velocities relative to the sample median and place a cut, above which an unusually large peculiar velocity is obtained.  In discussions of runaway stars, the peculiar space velocity threshold is generally placed at 30 km s$^{-1}$ \citep[e.g.][]{Hoogerwerf2001}.  A problem with this, in the present context, is that the considerable scatter in the measured radial velocities -- which may be due to binary reflex motion -- will propel some stars inappropriately into the high peculiar velocity group.  Accordingly, some stars that stand out on this criterion are less convincing candidate runaway stars.  Using $v_R,v_{\phi},v_Z =$ -11, 236, -0.6 km s$^{-1}$ as the median velocity components for the stars, defining the regional standard of rest within the 3--6 kpc distance range, we obtain 19 stars with a relative peculiar motion exceeding 30 km s$^{-1}$. Twelve of them are in common with the first selection.  Of the 22 unique objects found by both selections, eight are most divergent in $v$ which most closely aligns with radial velocity, whilst being only weakly or not at all anomalous in $u$ and $w$.   

Details of the superset of 22 stars that fall in either selection of outliers are given in Table~\ref{tab:extreme}.  Six of the stars listed were already picked out as likely runaways from Westerlund 2 by \cite[][see their Table 4]{Drew21}, and are identified as such in the final table column.  A sky map of where these stars are located and their relative proper motion vectors is presented in Fig.~\ref{fig:outlier_map}.  The 8 stars plotted in red are the stars with unexceptional $v_R$ and $v_Z$ that may prove to be spectroscopic binaries. Of the 14 stars, shown in blue, with outlier kinematics, six are the convincing Westerlund 2 ejections.  The cluster they are moving away from is visible as a grey blob in Fig.~\ref{fig:outlier_map} towards top right.  

\subsection{Westerlund 2 in its setting}
\label{sec:Wd2}

\begin{table*}
\caption{Data on the objects with the most anolmalous peculiar velocities according to the two criteria given in Section~\ref{sec:extreme}.  Columns 4 and 5 list $\log(T_{{\rm eff}})$ and $\log g$ obtained from blue AAOmega spectra that were reported by Mohr-Smith et al (2017).  For 3 stars in the table these parameters were not measured (spectra too noisy): the $\log(T_{{\rm eff}})$ values in brackets are less certain determinations from SED fitting described in the same paper. Column 6 gives the heliocentric radial velocity measured here using Pa12 and Pa13. 
The cited bounds on $D_{2.0}$, in column 7, correspond to the 16th and 84th percentiles of the EDSD inference.  Columns 8 to 10 contain the $v_R,v_{\phi},v_Z$ velocity components computed from the radial velocities, Gaia EDR3 proper motions and the column-7 distance. The entries in bold face pick out the components most implicated in the stars' outlier status.  The peculiar velocity in column 11 is obtained with reference to $(v_R,v_{\phi},v_Z) =$ (-11, 236, -0.6) km s$^{-1}$ as sample median and serving as the regional standard of rest. The last column provides codified comments such that 'Wd2' identifies the stars already seen as candidate runaways from Westerlund 2 (Drew et al 2018, 2021), 'SB?' identifies those stars that present only outlying radial velocities that may imply binary motion, while -- finally -- the two stars tagged with a '?' are of unclear status even as their peculiar velocities just pass the 30 km s$^{-1}$ threshold. 
}
{\centering
\begin{tabular}{ccccccccccll}
\hline
MMS & $\ell$ & $b$ &  $\log(T_{{\rm eff}})$ & $\log{g}$ & RV & $D_{2.0}$ & $v_R$ & $v_{\phi}$ & $v_Z$ $v_{pec}$ & Note\\
\# & \multicolumn{2}{c}{(degrees)} & & & (km s$^{-1}$) & (kpc) & \multicolumn{4}{c}{(km s$^{-1}$)} &  \\
\hline
  1236 & 284.2764 & -0.1635 & 4.622 & 3.98 & 16.2$\pm$8.8 & $4.90_{-0.31}^{+0.51}$ & -13$\pm$11 & 231$\pm$9 &   {\bf 18$\pm$1} &  19$\pm$3 [20] & Wd2 \\    
  1273 & 284.2989 & -0.5198 & 4.609 & 3.88 & 15.5$\pm$7.1 & $4.52_{-0.22}^{+0.32}$ & -17$\pm$8 & 228$\pm$8 &  {\bf -30$\pm$2} &  31$\pm$3 [28] & Wd2 \\      
  1308 & 284.3316 & -0.5836 & (4.402) & -- & -13.1$\pm$10.3 & $4.62_{-0.21}^{+0.29}$ & -28$\pm$7 & 255$\pm$10 & {\bf -39$\pm$3} &  46$\pm$5 [49] & Wd2 \\    
  1338 & 284.3784 & 0.0091  & 4.628 & 3.87 & -2.2$\pm$6.5 & $5.03_{-0.23}^{+0.32}$ & {\bf -32$\pm$7} &  244$\pm$6 & {\bf 40$\pm$2} &  47$\pm$4 [50] & Wd2 \\    
  1342 & 284.3865 & -0.9156 & 4.365 & 3.90 & -12.7$\pm$14.3 & $4.44_{-0.27}^{+0.42}$ & -15$\pm$11 & 257$\pm$14 & {\bf 16$\pm$1} &  27$\pm$11 &    \\    
  1353 & 284.4009 & -1.0804 & 4.428 & 3.85 & -27.0$\pm$2.6  & $4.29_{-0.23}^{+0.34}$ & -22$\pm$8 & {\bf 269$\pm$3} &   -1$\pm$1 &  35$\pm$4 & SB? \\      
  1354 & 284.4010 & -0.2890 & 4.551 & 4.09 & 8.6$\pm$9.2    & $4.73_{-0.35}^{+0.63}$ & {\bf -44$\pm$11} & 227$\pm$9 &  4$\pm$1 &  34$\pm$10 [34] & Wd2 \\    
  1356 & 284.4020 & -0.5481 & 4.591 & 4.23 & 94.9$\pm$23.7  & $4.00_{-0.18}^{+0.26}$ & {\bf -38$\pm$7} & {\bf 138$\pm$23} & {\bf -14$\pm$1} &  103$\pm$22 & \\    
  1374 & 284.4181 & -0.9274 & 4.587 & 4.19 & 19.8$\pm$11.1  & $4.91_{-0.29}^{+0.45}$ & -20$\pm$10 & 225$\pm$11 & {\bf -45$\pm$4} &  46$\pm$5 [44] & Wd2 \\     
  1379 & 284.4244 & -0.4512 & 4.363 & 3.97 & -38.7$\pm$14.1 & $4.63_{-0.31}^{+0.51}$ & {\bf -57$\pm$10} & {\bf 272$\pm$14} &  -10$\pm$2 &  59$\pm$11 &  \\   
  1382 & 284.4280 & -1.1354 & 4.339 & 3.35 & 70.3$\pm$4.3   & $4.90_{-0.26}^{+0.39}$ & 8$\pm$9 & {\bf 181$\pm$5}  &   1$\pm$0.4 &  58$\pm$6 & SB? \\     
  1440 & 284.5093 & -1.1386 & 4.330 & 3.96 & -18.5$\pm$7.0  & $5.49_{-0.34}^{+0.54}$ & -17$\pm$13 & {\bf 272$\pm$8} &  -2$\pm$1 &  36$\pm$8 & SB? \\    
  1728 & 284.8516 & -0.2267 & 4.300 & 3.20 & -18.5$\pm$2.1  & $4.95_{-0.44}^{+1.00}$ & -12$\pm$21 & {\bf 267$\pm$8} &   3$\pm$1 &  32$\pm$8 & SB? \\   
  1832 & 284.9472 & -0.5578 & 4.407 & 4.17 & -19.0$\pm$8.9  & $4.74_{-0.24}^{+0.35}$ & {\bf -46$\pm$7} & 254$\pm$9  &  -6$\pm$1 &  40$\pm$8 &  \\      
  1963 & 285.0503 & -0.6383 & 4.357 & 3.83 & 36.3$\pm$8.4   & $4.25_{-0.20}^{+0.29}$ & 6$\pm$8 & {\bf 209$\pm$8}  &  -6$\pm$1 &  32$\pm$8 & ? \\       
  2164 & 285.2164 & -1.1751 & 4.542 & 4.18 & 37.2$\pm$8.8   & $4.13_{-0.19}^{+0.26}$ & {\bf -33$\pm$5} & 198$\pm$9  & {\bf -22$\pm$2} &  49$\pm$7 & \\     
  2275 & 285.2978 & -0.3938 & (4.370) & -- & 42.2$\pm$8.9   & $5.35_{-0.47}^{+1.01}$ & -9$\pm$19 & 206$\pm$11  & {\bf -71$\pm$11} & 77$\pm$11 & \\   
  2342 & 285.3551 & -0.6633 & 4.433 & 3.88 & 44.2$\pm$5.7   & $5.31_{-0.34}^{+0.56}$ & -2$\pm$12 & {\bf 206$\pm$7}   &  -5$\pm$1 &  32$\pm$8 & SB? \\    
  2586 & 285.5667 & -1.4651 & 4.304 & 3.60 & 44.9$\pm$5.0   & $5.65_{-0.31}^{+0.48}$ & 3$\pm$11 & {\bf210$\pm$7}  &   6$\pm$0.3 &  30$\pm$8 & ? \\    
  2753 & 285.6841 & -0.3925 & 4.295 & 4.26 & 36.9$\pm$8.1   & $4.09_{-0.19}^{+0.27}$ & 0$\pm$7 & {\bf 205$\pm$8}  &  -1$\pm$1 &  34$\pm$8 & SB? \\   
  3190 & 285.9740 & -0.3766 & 4.329 & 3.37 & -1.8$\pm$4.0   & $3.63_{-0.16}^{+0.22}$ & {\bf 13$\pm$7} & 243$\pm$4 &  -0$\pm$1 &  25$\pm$7 & \\   
  3315 & 286.0606 & -0.9502 & (4.472) & -- & 47.2$\pm$9.7   & $5.74_{-0.39}^{+0.66}$ & {\bf 14$\pm$15} & 212$\pm$11 & -13$\pm$2 & 36$\pm$13 & \\
\hline
\end{tabular}
 }
\label{tab:extreme} 
\end{table*}

Work by \cite{Zeidler21} has made a detailed case for the systemic heliocentric radial velocity of Westerlund 2 being 15.9 km s$^{-1}$, with an error in the region of 1 km s$^{-1}$ (see their Fig. 3).  Both the gas associated with the cluster and the cluster's O star content appear to be consistent with this estimate.  It can be combined with the mean proper motion of the cluster to determine $v_R, v_{\phi}, v_Z$ for the cluster. For the proper motion and distance, we reuse the EDR3-based values given by \cite{Drew21}.  These are $\mu_{\ell,*} = -6.117\pm0.038$ mas yr$^{-1}$, $\mu_b = -0.437\pm0.030$ mas yr$^{-1}$ and $D_{2.0} = 4.45 (+0.15,-0.14)$ kpc. These data convert to $(v_R,v_{\phi},v_Z) = (-10.5\pm4.2,228.7\pm1.6,-2.0\pm0.7)$ km s$^{-1}$. 

We now compute the ejection speeds for the six runaway candidates relative to Westerlund 2, using the above $v_R,v_{\phi},v_Z$ estimates. The revised estimates, compared to the listed peculiar velocities in Table~~\ref{tab:extreme}, differ at the level of up to 3 km s$^{-1}$.  However, as they are to be preferred in discussions of Westerlund 2 and its OB-star diaspora, they are provided as second values in square brackets in the table.  \cite{Drew2018} estimated central values of 32 and 47 km s$^{-1}$ for the ejection speeds of \#1273 and \#1338 respectively: these are now 28$\pm$3 and 50$\pm$4 km s$^{-1}$.  

It is worthy of note that another star of the six, \#1308, is now known as WR20aa \citep{RomanLopes2016} and is listed in the online WR catalogue\footnote{http://pacrowther.staff.shef.ac.uk/WRcat/index.php} \citep{WRcat}.  At bluer wavelengths, this star presents both emission and absorption lines in its spectrum, but only absorption is apparent in the high Paschen series lines of the CaT region.

\section{Discussion}
\label{sec:discussion}

\subsection{High peculiar-velocity stars}
\label{sec:high_vel}

Eight of the list of kinematic outliers (relative to the regional standard of rest defined in Section~\ref{sec:extreme}) in Table~\ref{tab:extreme} are neither ejections from Westerlund 2 nor likely interloping spectroscopic binaries.  We cannot identify a likely point of origin for any of them.  These stars warrant brief comment based on what we do know about them, so far.

\begin{figure}
\begin{center}
\includegraphics[angle=-90,width=1.0\columnwidth]{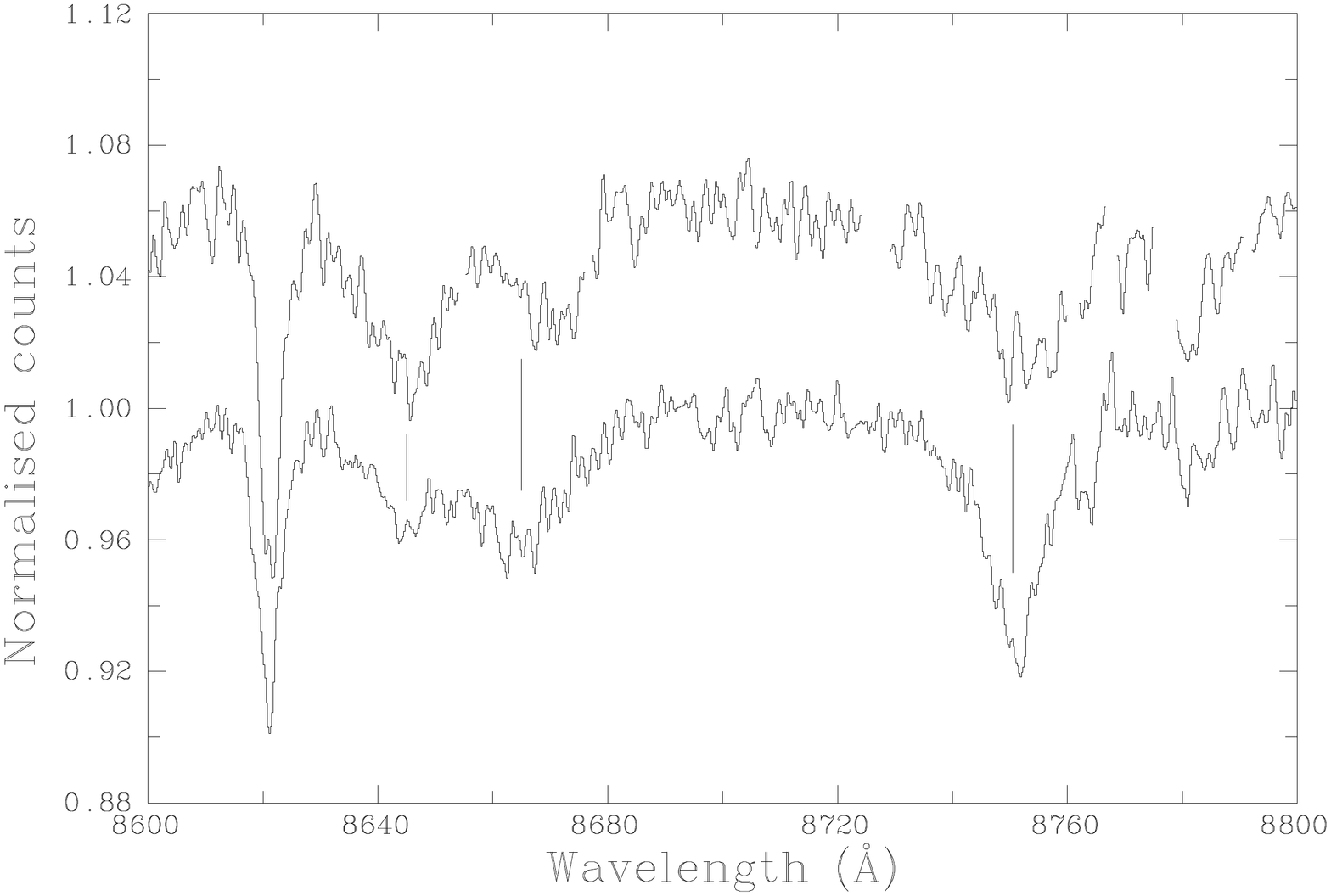}
\caption{Part of the 1700D spectra of \#1356 (top and displaced by +0.06) and \#1338 (bottom), including the Pa 12 and Pa 13 features.  The vertical line segments between the two spectra mark the broader weaker DIB, and the rest wavelengths for Pa 13 and Pa 12 (working from left to right). This is to show how \#1356 presents a notable positive velocity shift (measured as +109 km s$^{-1}$, telescope frame), with reference to the rest wavelengths and \#1338, where the shift is small (+11 km s$^{-1}$, observer frame).  Both DIBs in the plotted range, line up well.   
}
\label{fig:1356}
\end{center}
\end{figure} 

VPHAS-OB1-1356:- This object has the most extreme peculiar velocity of the set at 103$\pm$22 km s$^{-1}$.  Much -- but not all -- of this originates in the radial velocity, which carries the highest uncertainty of the sample due to the low contrast of its photospheric absorption (see Fig.~\ref{fig:1356} -- $v\sin i \simeq 310$ km s$^{-1}$, MS17).  Its parallax is one of the more secure ($\pi/\sigma_{pi} = 16.04$, placing the star $\sim$4 kpc away (see Table~\ref{tab:extreme}).  MS17 determined the extinction towards this star to be $A_0 = 5.68 (+0.05,$-$0.06$) mag.  These data, combined with the star's apparent magnitude ($r = 13.74$, also from MS17), lead to an estimated absolute visual magnitude 0f $-4.4$~mag.  We assume in this calculation that the intrinsic colour, $V - r$ is $-0.3$, and that $A_r/A_0 = 0.85$.  According to the least-squares trends presented by \cite{Martins2005}, this absolute magnitude corresponds to an O8 dwarf.  However, the 39 kK effective temperature of this star and $\log g$ of 4.23, obtained from the fits to blue-wavelength line profiles by MS17, point to a more luminous O5.5 -- O6 dwarf ($M_V \sim -5$), on the Martins et al scales.  This discrepancy falls within the scatter associated with these trends. That both the effective temperature and surface gravity estimates are higher than the Martins et al mean dwarf values may have arisen from a parameter degeneracy in the spectroscopic fitting.  Even if the measured radial velocity is affected by binary motion, the high relative proper motion of the star alone is enough to give an in-sky peculiar velocity relative to the region mean of 49 km s$^{-1}$ -- well in excess of the usual 30 km s$^{-1}$ threshold.  It is presently roughly 16 arcmin from the centre of Westerlund 2 on the sky, and most likely has never been much closer.

VPHAS-OB1-2164:- This object is the other clear O dwarf among the set of 10.  The fit to blue spectroscopy (MS17) gave an effective temperature of 35 kK, and $\log g = 4.18$, indicating O8V \citep{Martins2005}.  As for \#1356, estimation of its absolute magnitude using its parallax, photometrically-determined extinction and observed apparent magnitude, favours a later type (O9V).   Its trajectory in the plane of the sky, if linear, passed Westerlund 2 at a minimum distance of $\sim 8$ arcmin, over 10 Myr ago.  Accordingly, the cluster is an unlikely point of origin.    

VPHAS-OB1-3190:- This star only deviates significantly from the region mean motion in its $v_R$ component, a property that originates in its relative longitudinal proper motion being significantly more negative than the mean.  Because of this, \#3190 stands out in Figure~\ref{fig:outlier_map} as the only star moving directly towards decreasing longitude. It could do this if it were in the foreground of the sample, which -- to an extent -- it is, at a distance $D_{2.0} \simeq 3.6 kpc$.  However, the 14 stars in the sample at the same or a shorter distance present longitudinal proper motions at least 1.4 mas yr$^{-1}$ less negative.  Its distance, extinction and apparent magnitude point to $M_V \simeq -3.5$~mag. Broadly, this reconciles with the spectroscopically-determined effective temperature of 21 kK, and surface gravity ($\log g \simeq 3.4$) for the star, as a B2 giant\footnote{see Schmidt-Kaler's pages in the 1982 Landolt-B\"ornstein tables, https://ui.adsabs.harvard.edu/abs/1982lbg6.conf.....A/abstract}.  The age in such a case exceeds 10 Myr \citep[e.g.][]{Ekstrom2012}, opening up the possibility that \#3190 originated $\sim4^{\circ}$ or more away from its present sky position.

VPHAS-OB1-1379:-  This star is an early B-type ($\sim$ B1) dwarf.  The current trajectory of \#1379 traces back to a minimum separation from Westerlund 2 of just $\sim$5 arcmin.  It is a rapid rotator, with $v\sin i \sim 310$ km s$^{-1}$.  It is a small amount brighter than $g = 16$~mag, resulting in its spectrum being one of the least well exposed in the set (section~\ref{sec:spectroscopy}):.  These factors conspire to produce the fifth-highest radial velocity error (14.1 km s$^{-1}$) for this star.  Like \#1356, there is no room for doubting it has a high peculiar velocity, given that its longitudinal proper motion alone is sufficient qualification. 

VPHAS-OB1-1342, -1832:-  Again, the fits to the blue spectroscopy identify both stars as early B-type.   \#1342, has an upwards trajectory through the Galactic plane -- heading towards Westerlund 2.  The trajectory of \#1832 is almost collinear, in the plane of the sky, with that of \#1379 (see Figure~\ref{fig:outlier_map}, but the respective patterns of their space-velocity components indicate 3D divergence.  

\subsection{On the impact of binary motion}
\label{sec:binary_motion}

\begin{figure}
\begin{center}
\includegraphics[width=0.8\columnwidth]{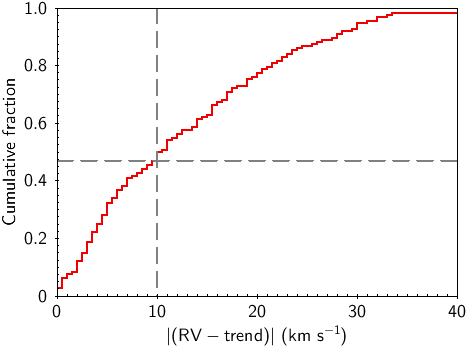}
\caption{Cumulative function of the absolute difference between measured heliocentric velocity and the mean regression ($-25.1124 + 7.3325*D_{2.0}$, black dashed line plotted in Figure 6). An absolute velocity difference of up to 10 km s$^{-1}$ matches to a cumulative fraction of 47\% (53 \% exceeding, dashed horizontal and vertical lines). 
}
\label{fig:cf}
\end{center}
\end{figure} 

Only for two stars in the sample of 150 stars between 3 and 6 kpc away is it already clear that there is no marked binary motion.  These are \#1273 and \#1338, that were observed by \cite{Drew2018}, more than once and at time intervals designed to pick up significant reflex motion.  No such motion was detected above a few km s$^{-1}$ \citep[see Section 4.2 in][]{Drew2018}. For the rest, further spectroscopy will be needed.

\cite{Sana2013} collected repeat spectroscopy of 360 O stars in the Tarantula Nebula to build a statistical sample of radial velocity changes. They found that 0.35$\pm$0.03 of their sample exhibited radial velocity shifts exceeding 20 km s$^{-1}$.  From this they deduced a binary fraction of 0.51$\pm$0.04.  As a coarse comparison with this result, the cumulative function of the absolute difference between the measured heliocentric radial velocity and the linear regression of Figure~\ref{fig:rv_v_distance} is shown in Figure~\ref{fig:cf}.  The absolute difference exceeds 10 km s$^{-1}$ for 53 \% of the sample -- an appreciably larger fraction than 35 \%.  Whilst the $\pm$10 km s$^{-1}$ boundary is not identical with Sana et al.'s criterion, the presence of an excess in the distribution due to peculiar space motion would seem to be permitted.  A final point of logic to note is that moderate binary motion can act to reduce as well as enhance a radial space velocity anomaly.   

\subsection{The OB population 3D kinematics}
\label{sec:OB_kinematics}

\begin{figure}
\begin{center}
\includegraphics[width=0.8\columnwidth]{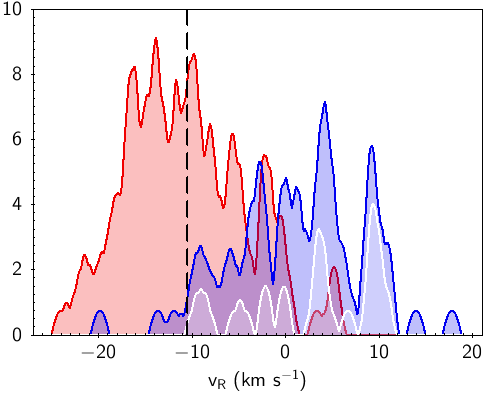}
\caption{Comparison of the distribution of $v_r$ as obtained from refined OB Carina spectroscopic sample (red) with the distribution for younger open clusters obtained by Tarricq et al (2021).  They are presented as KDE plots with an arbitrary and matching kernel width.  The Carina data for 125 stars is in red.  Open clusters with estimated ages of up 100 Myr are in blue (79 clusters).  A subset of 24 with estimated ages of less than 20 Myrs are superposed in white. These clusters are mainly within one kpc of the Sun. The vertical black dashed line marks $v_r = -10.5$ km s$^{-1}$, the mean obtained for Westerlund 2.  Whilst the two distributions overlap, the Carina OB stars along with Westerlund 2 distinctively favour infall. 
}   
\label{fig:U_comparison}
\end{center}
\end{figure} 

To set a context, we compare the average $v_R, v_{\phi}, v_Z$ velocity components obtained from the Carina OB-star sample and for Westerlund 2 with the results on 1382 open clusters (OCs), located mainly in the solar neighbourhood, that were presented by \cite{Tarricq21}. The advantage of making a comparison with OC kinematics is that the clusters can be aged, making it possible to separate out a suitably young subset.  \cite{Tarricq21} compiled cluster space velocities using Gaia DR2 astrometry in combination with radial velocity data from Gaia's RVS and a range of ground-based spectroscopic surveys.  In turn, we use the data flagged by them as 'high quality' and limit our attention to OCs with log[age(yr)] $< 8.0$ at Galactocentric radii in the range 7.8--8.8 kpc (a sample of 79 OCs).  From these we obtain ($\bar{v_R},\bar{v_{\phi}},\bar{v_Z}) = (1.1, 238.5, -0.8)$ km s$^{-1}$. The dispersions that go with these means are $(6.9, 6.2, 4.0)$ km s$^{-1}$. After requiring $3 < D_{2.0} < 6$ kpc, $\pi/\sigma_{\pi} > 5$ and excluding the obvious outliers (leaving 125 OB stars), the analogous data for the Carina sample, representing its regional standard of rest, would be $(-10.6, 236.3, -0.5)$ km s$^{-1}$, with dispersions $(6.3, 14.9, 4.1)$ km s$^{-1}$.  

The mean azimuthal and vertical components of motion ($v_{\phi}$ and $v_Z$) for the Carina OB stars are entirely compatible with the younger OCs in the solar neighbourhood.  But there is a difference in $\bar{v_R}$ at the level of around twice the dispersion in either sample.  The sense of the difference is that the Carina OB stars clearly favour an inward drift towards the Galactic Centre.  To illustrate this, Figure~\ref{fig:U_comparison} superimposes the two distributions.  Shrinking the OC sample down to 24 clusters younger than 20 Myr (also plotted) does not reduce the offset.  The likely age of the bulk of the Carina OB sample is in the region of a few to 10 Myr, making it plausible that the bulk kinematics still carry the imprint of the molecular clouds they were born in.  This leads to the inference that the star-forming gas was itself infalling with $v_R \sim 10$ km s$^{-1}$.  The other comparison to make here, accordingly, is with results obtained for high-mass star-forming regions. These are provided for the northern hemisphere, mainly, on the basis of collected VLBI maser data by \cite{Reid2019}.  In modelling their data, their best-performing result yields a mean inward drift of 6.0$\pm$1.4 km s$^{-1}$ (here, the standard error, rather than the dispersion, is given -- see their Table 3).

It was shown in Section~\ref{sec:Wd2} that the space velocity of the massive open cluster, Westerlund 2, in component form is $(-10.5,228.7,-2.0)$ km s$^{-1}$.  The value of the $v_R$ term is marked by a black-dashed vertical line in Fig.~\ref{fig:U_comparison}, to show how it sits right at the heart of the reduced OB sample of 125 stars.  In contrast, there is a potentially significant offset in $\bar{v_{\phi}}$ ($-$7.6 km s$^{-1}$ and $-$9.8 km s$^{-1}$ with respect to the Carina OB stars and the young \cite{Tarricq21} OCs).  The standard errors on $\bar{v_{\phi}}$ for these two samples are respectively 0.6 and 0.7 km s$^{-1}$, suggesting that Westerlund 2's azimuthal motion may lag general disk rotation by a few km s$^{-1}$.  \cite{Reid2019} obtained a mean lag of 4.3$\pm$5.6 km s$^{-1}$ in their modelling of maser kinematics, relative to a circular speed, of 236 km s$^{-1}$.  These figures line up quite well with the Westerlund 2 $v_{\phi}$ result.  But there are other proposed rotation laws in the literature: for example, the \cite{Eilers19} law, based on luminous red giant data, would specify a circular speed within a few tenths of 229 km s$^{-1}$.  If this is adopted instead as the reference, $v_{\phi}$ for Westerlund 2 matches to within the errors, while the mean rotation for the OB stars and OCs are then faster by 7.3 and 9.6 km s$^{-1}$ respectively.  The comparisons with the gas and young stellar tracers are likely to be the better choices here, but it cannot yet be claimed that all the uncertainties and subtleties in defining the rotation law have been tied down.

An important point emerging from the above is that the measured OB-star RV -- dominating the values of $v_{\phi}$ obtained -- yield credible results. With binary motions folded in, they are going to be more dispersed than they would have been without them.   Indeed, the usual pattern that $\sigma_{v_R} > \sigma_{v_{\phi}} > \sigma_{v_Z}$ \citep[e.g.][]{DehnenBinney1998} is not followed because of the inflation of $\sigma_{v_{\phi}}$ (middle panel, Figure~\ref{fig:rv_v_distance}).  If the intrinsic dispersion of 6.2 km s$^{-1}$ is carried across from the OC sample and we adopt a measurement error dispersion of 7 km s$^{-1}$, the remaining contribution to the quadrature sum that can be assigned to binary motion is in the region of 12 km s$^{-1}$.

The pattern of vertical motions ($v_Z$) present in the data are as expected.  We find that our results are consistent to within the uncertainties with $w_{\odot} \simeq 7$ km s$^{-1}$, and observe a scatter compatible with previous results for young, blue stars from \cite{Tarricq21} back to \cite{Aumer2009} and \cite{DehnenBinney1998}.  






\section{Conclusions}
\label{sec:conclusions}

We have explored the measurement of OB-star radial velocities using Paschen lines in the far red, adjacent to the Calcium triplet lines. The sample studied consists of 164 OB stars, in the far Carina Arm at distances mostly between 3 and 6 kpc, spread across a 2-degree diameter field. This is a pilot for using the red end of the optical spectrum to obtain useful measurements on significantly reddened, distant massive stars that are challenging to access at blue wavelengths.  It has been shown it can be done and that credible resultant space motions are obtained: in particular, the azimuthal $v_{\phi}$ components that are almost aligned with the line of sight are found to be distributed around a plausible mean circular speed of $\sim$236 km s$^{-1}$. 

Imperfect sky subtraction will usually combine with the weakness and broadening of these lines to limit the precision of RV determination.  We have found it leads to errors typically in the range 5--10 km s$^{-1}$.  The results are based on measuring the cores, only, of the Pa 12 and Pa 13 lines -- leaving out the pressure-broadened wings.  A way forward here, that can be explored with the forthcoming pan-optical wavelength coverage of the new WEAVE and 4MOST spectrographs, would be to bring in additional longer-wavelength Paschen lines that are clear of blending with the Calcium triplet lines. The recent Gaia DR3 data release has shown that the RVS on board the Gaia mission are not suited to significantly reddened OB stars (Appendix~\ref{sec:Gaia_spectra}).

We have combined the AAOmega RV data with Gaia astrometry in order to expose the 3D kinematics of the spectroscopic OB sample.  This has provided a list of six runaway stars with full space velocities that have been identified in previous literature as likely ejections from Westerlund 2 (see Table~\ref{tab:extreme} and Fig.~\ref{fig:outlier_map}). Five of them -- stars \#1236, \#1273, \#1338, \#1354 and \#1374 -- are O-type \citep{MMS2017, Drew2018}, while \#1308 has the unusual extreme spectral type O2If*/WN6 \citep[also known as WR 20aa, see][]{RomanLopes2011}. In addition, eight high peculiar velocity stars are identified that are not ejections from Westerlund 2. In eight more cases it is not yet possible to determine if the outlier radial velocity is due to binary reflex or high systemic motion. 

The 3D kinematics have also revealed that the bulk azimuthal motion of the OB-star sample is in line with mean Galactic disk rotation, while, in the radial direction, there is the feature of a modest, but convincing, net infall ($v_R \sim -10$ km s$^{-1}$).  The same infall is exhibited by Westerlund 2, suggesting a total picture in which the giant molecular cloud complex giving rise to this young generation of stars was infalling at this speed, on average. The section of the Carina Arm investigated lies mainly between 3 and 6 kpc away, a couple of degrees inside the tangent longitude, affording us a tidy nearly end-on column of OB stars.  An early priority for the future should be to roll out an expanded programme of medium-resolution spectroscopy of the wider far Carina Arm, to take in many of the \cite{MMS2017} OB-star catalogue -- especially as it is already known from proper motion data \citep{Drew21} that there is as-yet unexplained complexity in the massive star kinematics.  Here we have examined only a small part of it.

Both the far-red spectra (including the four emission line objects in Table 2) and the measured radial velocities plus other parameters are provided as supplementary material.  A specification of the table of parameters and radial velocities is given in the Appendix.

\section*{Acknowledgements}


 The spectra this paper is based on were obtained via a service programme executed on the Australian Astronomical Telescope. 
The origin of the sample lies in data products from the ESO Telescopes at the La Silla Paranal Observatory -- specifically, the VST Photometric H$\alpha$ Survey of the Southern Galactic Plane and Bulge (VPHAS+, ESO programme 177.D-3023). 
Data from the European Space Agency mission Gaia (https://www.cosmos.esa.int/gaia), processed by the Gaia Data Processing and Analysis Consortium (DPAC, https://www.cosmos.esa.int/web/gaia/dpac/consortium) have also been used. 
The analysis presented was performed using {\sc TopCat} \citep{Taylor2005} alongside {\sc DIPSO} and its {\sc ELF} gaussian fitting tool developed respectively by I. Howarth and P. Storey.  

The radial velocity measurements reported here began as FBP's masters project at University College London under JED's supervision: M. J. Barlow's support as second supervisor is gratefully acknowledged. P. Storey is thanked for advising on the ELF method of error calculation.  MMS set up the 2014 AAOmega observations and carried out the data reduction whilst a PhD student at the University of Hertfordshire, funded by the Science \& Technology Facilities Council of the UK.  JED made the final full set of measurements and wrote the paper.  

An anonymous referee is thanked for their helpful comments.



\section*{Data Availability}

The main data products associated with this paper are directly available as a zipped file of far-red spectra and as a table of measurements, both supplied as supplementary material (see Appendix~\ref{sec:full_list}).  


\bibliographystyle{mnras}
\bibliography{JED_papers}

\appendix


\section{Description of supplementary materials}
\label{sec:full_list}

Parameters and measurements of all 164 stars included in the sample presented in this paper are available as a machine-readable table, OBstar\_parameter\_table.fits provided as supplementary material.  The columnns in this table are identified below in Table~\ref{tab:TableAp}.

The spectra themselves, in fits format, are contained within the  file, collected\_OBspectra.zip, also provided as supplementary material.  The names of the individual spectra are the serial numbers NNNNN, taken from the full \cite{MMS2017} catalogue names of form VPHAS-OB1-NNNNN.  They are presented as wavelength-calibrated by the extraction pipeline (without heliocentric correction).

\label{lastpage}

\newpage
\onecolumn
\begin{longtable}[h!]{llcl}
\caption{Contents of FITS table in supplementary materials.  The naming, position, $g$ magnitude and stellar parameter estimates in columns 1 to 25 are carried across from Mohr-Smith et al 2017. The radial velocity measurements based on the Paschen 12 and 13 lines are in columns 26--29.  The most directly relevant Gaia EDR3 parameters are restated in columns 30--37.  Proper motions converted into Galactic coordinates, distance estimates (obtained via the EDSD method of Luri et al. 2018), and the derived space velocity components occupy columns 38 to 50.
}
\\\label{tab:TableAp}
No & Column name & Units & Description \\ \hline
1 & Star name & None &  The serial number, NNNNN, rising with increasing Galactic longitude, extracted from\\
 & & & the MS17 catalogue name.  The full form of the name is VPHAS-OB1-NNNNN. \\
2 & RA\_2000 & degrees &  Right Ascension J2000 (VPHAS$+$)\\ 
3 & DEC\_2000 & degrees &  Declination J2000 (VPHAS$+$)\\ 
4 & GLON & degrees &  IAU 1958 Galactic longitude \\ 
5 & GLAT & degrees &  IAU 1958 Galactic latitude \\ 
6 & g & mag &  VPHAS$+$ g band magnitude \\ 
7 & e\_g & mag &  random error on VPHAS$+$ g-band 
magnitude \\
8 & log(Teff K) & None &  Estimated effective temperature from photometric fits (median of posterior) \\ 
9 & eu\_logTeff & None &  Upper uncertainty on logTeff (84th percentile of posterior) \\ 
10 & el\_logTeff & None &  Lower uncertainty on logTeff (16th percentile of posterior) \\ 
11 & A0 & mag &  Estimated extinction, A0, from photometric fits (median of posterior) \\ 
12 & eu\_A0 & mag &  Upper uncertainty on A0 (84th percentile of posterior) \\ 
13 & el\_A0 & mag &  Lower uncertainty on A0 (16th percentile of posterior) \\ 
14 & Rv & None &  Photometric estimate of extinction law parameter, $R_V$ \\ 
15 & eu\_Rv & None &  Upper uncertainty on RV (84th percentile of posterior) \\ 
16 & el\_Rv & None &  Upper uncertainty on RV (16th percentile of posterior) \\ 
17 & speclog(Teff K) & None & Spectroscopic estimate of effective temperature (median of posterior) \\
18 & eu\_speclogTeff & None & Upper uncertainty on speclogTeff (84th percentile of posterior) \\
19 & el\_speclogTeff & None & Lower uncertainty on speclogTeff (84th percentile of posterior) \\
20 & logg & None & Spectroscopic estimate of log(g cm s$^{-2}$) \\
21 & eu\_logg & None & Upper uncertainty on logg (84th percentile of posterior) \\
22 & el\_logg & None & Lower uncertainty on logg (16th percentile of posterior) \\
23 & vsini & km s$^{-1}$ & Spectroscopic estimate of projected rotation speed \\
24 & eu\_vsini & km s$^{-1}$ & Upper uncertainty on vsini (84th percentile) \\
25 & el\_vsini & km s$^{-1}$ & Lower uncertainty on vsini (16th percentile) \\
26 & Pa\_lambda & \AA\ & Measured Pa 12 wavelength (from combined Pa 12 + Pa 13 fit) \\
27 & e\_Pa\_lambda & \AA\ & Fit uncertainty in measured wavelength \\
28 & Pa\_vel\_hc & km s$^{-1}$ & Heliocentric radial velocity (from Pa\_lambda) \\
29 & e\_Pa\_vel\_hc & km s$^{-1}$ & Error on heliocentric radial velocity \\
30 & EDR3Name & None &  Gaia EDR3 source identifier \\
31 & Plx & mas &  Gaia EDR3 parallax, $\pi$ \\ 
32 & e\_Plx & mas &  Uncertainty in parallax, $\sigma_{\pi}$ \\
33 & RPlx & None & parallax/uncertainty, $\pi/\sigma_{\pi}$ \\ 
34 & pmRA & mas yr$^{-1}$ &  Proper motion in right ascension\\  
35 & e\_pmRA & mas yr$^{-1}$ &  Uncertainty in proper motion in right ascension \\ 
36 & pmDE & mas yr$^{-1}$ &  Proper motion in declination\\ 
37 & e\_pmDE & mas yr$^{-1}$ &  Uncertainty in  proper motion in declination \\ 
38 & pml & mas yr$^{-1}$ &  Longitudinal proper motion \\ 
39 & e\_pml & mas yr$^{-1}$ &  Uncertainty in longitudinal proper motion \\ 
40 & pmb & mas yr$^{-1}$ &  Latitudinal proper motion \\ 
41 & e\_pmb & mas yr$^{-1}$ &  Uncertainty in latitudinal proper motion \\
42 & D\_2.0 & kpc &  Distance estimate (EDSD inversion: offset 0.030 mas, length scale 2.0 kpc) \\
43 & eu\_D\_2.0 & kpc &  Upper uncertainty on distance (to 84th percentile of posterior) \\ 
44 & el\_D\_2.0 & kpc &  Lower uncertainty on distance (to 16th percentile of posterior) \\ 
45 & v\_R & km s$^{-1}$ & Galactocentric radial velocity \\
46 & e\_v\_R & km s$^{-1}$ & Error on v\_R \\
47 & v\_phi & km s$^{-1}$ & Azimuthal velocity ($+$ve in direction of disk rotation) \\
48 & e\_v\_phi & km s$^{-1}$ & Error on v\_phi \\
49 & v\_Z & km s$^{-1}$ & Velocity out of the Galactic plane \\
50 & e\_v\_Z & km s$^{-1}$ & Error on v\_Z \\
\hline

\end{longtable}

\twocolumn
\section{Cross-match with Gaia spectra}
\label{sec:Gaia_spectra}

\begin{figure}
\begin{center}
\includegraphics[width=1.0\columnwidth]{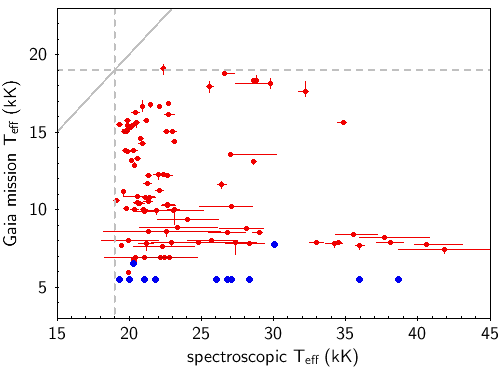}
\caption{
Gaia mission stellar effective temperatures from the BP/RP prism spectra (93 red data points) and as adopted for the radial velocity measurement (12 blue points) as a function of the effective temperatures determined by MS17 from blue AAOmega spectra. The solid grey line is the equality line, while the vertical and horizontal dashed lines mark 19 kK, the minimum $T_{{\rm eff}}$ imposed on the sample.  These spectra, with their low-contrast absorption lines, are usually misconstrued as lower-extinction cool stars by the Gaia pipeline.
}
\label{fig:carina_teffs}
\end{center}
\end{figure}

As this paper was being revised, the Gaia DR3 release occurred.  This made available stellar parameter determinations from the low resolution prism data and extended the database of radial velocities derived from RVS observations.  A fresh cross-match was performed in order to determine whether any of the new radial velocities would be helpful to the present study. This revealed that 12 of the sample of 164 stars now have Gaia radial velocities.  But it turns out that there is reason to distrust them.  This is most easily and succinctly illustrated by the comparison between ground-based and Gaia stellar effective temperatures in Fig.~\ref{fig:carina_teffs}: there is an evident failure to recognise the sample stars for the O and early B stars that MS17 confirmed them to be. 

The Gaia pipelines have been set up to give good results on the cool stars that dominate in any stellar population.  The problem that OB stars present for analysis has two parts to it.  The first part is the well-known degeneracy between extinction and intrinsic colour (i.e. spectral type) that means reddened OB stars are extremely challenging to separate out from cool less-reddened stars by means of spectral energy distribution analysis, such as that performed on Gaia prism data.  There is a discussion of this problem in the extinction mapping work of \cite{Sale2009}.  The low Gaia-assigned effective temperatures in Fig.~\ref{fig:carina_teffs} are accompanied by extinctions that are typically $\sim$25\% lower than those of MS17.  The second issue follows from the low contrast of the photospheric absorption against the continuum in OB spectra: it greatly inhibits the discrimination that might otherwise correct the low assigned effective temperatures -- leaving open to the pipeline the option of favouring parameter sets that combine low surface gravities and metallicities with cool temperatures.  As a result, the parameters reported for 93 stars (of the sample of 164) in DR3 unduly favour metal-poor high-luminosity cool stars.  For 2 of the 12 stars measured for radial velocity the metallicity problem is especially acute: remarkably, they are assigned [Fe/H] = $-$2.5, $-$3.0.  The other 10 are given default Sun-like stellar parameters.

\begin{figure}
\begin{center}
\includegraphics[width=0.8\columnwidth]{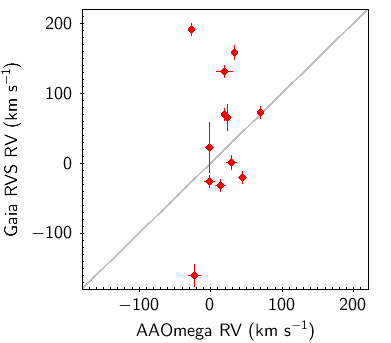}
\caption{
Gaia DR3 radial velocities compared with the values obtained in this study (12 stars). The solid grey line is the equality line.  The correlation coefficient computed relative to the best-fit linear regression through these data points is only 0.18.
}
\label{fig:carina_RVs}
\end{center}
\end{figure}

A comparison of the radial velocities for the 12 stars with Gaia measurements is shown as Fig.~\ref{fig:carina_RVs}.  The much greater spread of the Gaia results is striking.  Given the problems just described, and the lack of correlation, it does not yet seem safe to assume they are reliable. 

\end{document}